\documentclass[prc,aps,amsfonts,twocolumn,floatfix,superscriptaddress,nofootinbib]{revtex4-1}

\usepackage{lineno,hyperref}
\modulolinenumbers[5]
\usepackage{graphicx,amssymb,color,amsmath}
\usepackage{natbib}
\usepackage{braket}
\usepackage{cases}
\usepackage{color}
\usepackage{mathtools}
\usepackage{caption}
\usepackage{subcaption}
\usepackage{multirow}

\definecolor{orange}{rgb}{1,0.5,0}

\begin{document}

\title{Coupling charge-exchange vibrations to nucleons in a relativistic framework: \\ effect on Gamow-Teller transitions and beta-decay half-lives}

\author{Caroline Robin}
\email{carolr8@uw.edu}
\affiliation{Institute for Nuclear Theory, University of Washington, Seattle, WA 98195, USA.}
\affiliation{JINA-CEE, Michigan State University, East Lansing, MI 48824, USA.}

\author{Elena Litvinova}
\affiliation{Department of Physics, Western Michigan University, Kalamazoo, MI 49008-5252, USA}
\affiliation{National Superconducting Cyclotron Laboratory, Michigan State University, East Lansing, MI 48824, USA}

\date{\today}

\begin{abstract}
The nuclear response theory for isospin-transfer modes in the relativistic particle-vibration coupling framework is extended to include coupling of single nucleons to isospin-flip (charge-exchange) phonons, in addition to the usual neutral vibrations. 
This new coupling introduces dynamical pion and rho-meson exchange, beyond the Hartree-Fock approximation, up to infinite order.
We investigate the impact of this new mechanism on the Gamow-Teller response of a few doubly-magic neutron-rich nuclei, namely $^{48}$Ca, $^{78}$Ni, $^{132}$Sn and $^{208}$Pb.
It is found that the coupling to isospin-flip vibrations can have a non negligible impact on the strength distribution and quenching of the giant resonance, globally improving the agreement with the experimental data. 
The corresponding beta-decay half-lives of $^{78}$Ni and $^{132}$Sn are also calculated, and found to be decreased by a factor $\sim 2$ by the inclusion of the new phonons. 
\end{abstract}

\maketitle

\section{Introduction}
\begin{figure}
\centering{\includegraphics[width=\columnwidth] {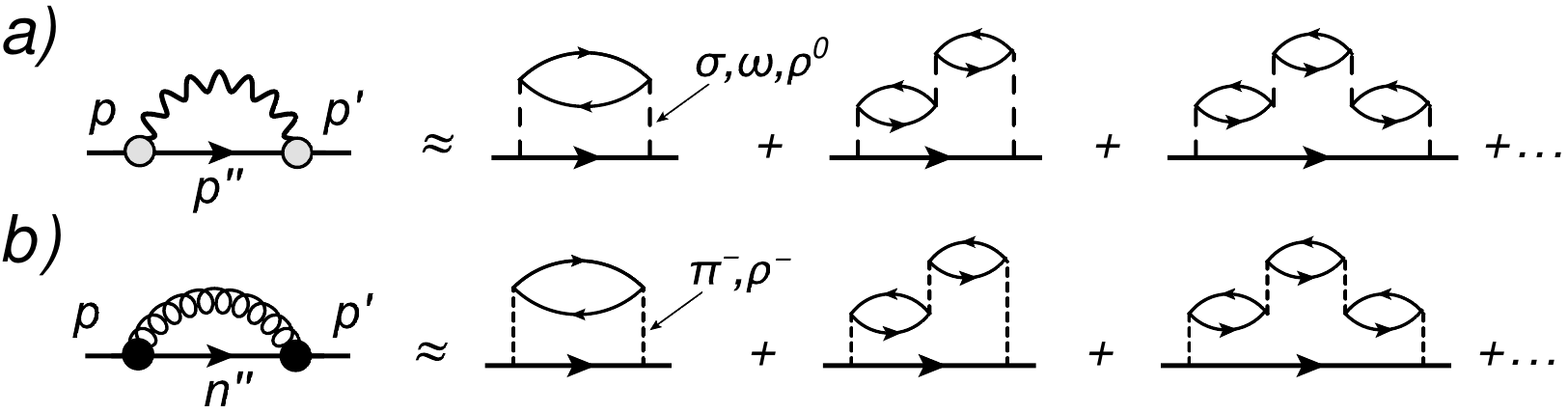}}  \hfill % 
\caption{Coupling of nucleons to neutral (a) and charge-exchange (b) phonons.}
 \label{f:exp-phon} 
\end{figure}

A consistent treatment of single-particle and collective degrees of freedom in nuclei remains one of the central challenges in modern nuclear structure theory.
Starting from an {\it ab-initio} G-matrix derived from a bare nucleon-nucleon interaction \cite{Brueckner1954,BHF1974,Shen2016}, or from a phenomenological parametrization of a density functional \cite{Skyrme,GognyD1,RING1996}, one can reproduce relatively well bulk properties of a wide range of nuclei within mean-field theories. 
It is well known, however, that this level of approximation fails to reproduce single-particle spectra around the Fermi level, due to neglected retardation effects in the one-nucleon self-energy.
Moreover, the associated theory for the response of nuclei to external fields, known as random-phase approximation (RPA), can only provide a poorly detailed description of nuclear excitations.
To remedy these deficiencies one must consider higher-order dynamical processes in the nucleonic self-energy. Such corrections arise from strong medium polarization effects described as
virtual excitations of particle-hole (p-h) pairs. 
The particle-vibration coupling (PVC) scheme offers a way to include such processes up to infinite orders, by considering the coupling of single nucleons to collective vibrations of the nucleus, that are built of coherent interacting p-h excitations.
Historically this framework was inspired from the pioneering idea of Bohr and Mottelson \cite{BohrMottelson}, and has been developed and applied in different contexts over the years. Non-relativistic versions include the nuclear field theory \cite{BORTIGNON1977,BES1976}, extensions of the Landau-Migdal theory \cite{LitvinovaTselyaev2007,TselyaevSpeth2007,KAMERDZHIEV20041} and quasiparticle-phonon model (see \textit{e.g.} Ref. \cite{soloviev1992} and references therein). More recently self-consistent implementations of the PVC scheme have emerged, and have been applied to both single-particle motion and two-body response in different channels, 
in closed and open-shell nuclei. These include the non-relativistic PVC based on Skyrme interaction \cite{Colo2010,Cao2014,Niu2012,Niu2016,NIU2018} and the relativistic one \cite{Litvinova2006,Litvinova2007,Litvinova2008,Marketin2012,Litvinova2014,Robin2016}.
All of the aforementioned studies, however, have restricted the space of vibrations that enter the PVC mechanism, to neutral (non isospin-flip) excitations, which, in the covariant case based on relativistic mean field (RMF), resum scattering of $\sigma$, $\omega$ and $\rho$ mesons on nucleonic p-h pairs, as illustrated in Fig. \ref{f:exp-phon}-a). 
The justification was that most low-energy collective modes are surface vibrations of neutral nature which consequently should give the dominant contribution to the PVC mechanism.
It has been early realized, however, that isospin-flip modes of collective character can also occur at low-energy \cite{MIGDAL1990}. These modes, which were thought to be related to the onset of pion condensation, can therefore potentially couple to single nucleons.
The influence of charge-exchange (CE) phonons on the single-particle shell structure of $^{100}$Sn and  $^{132}$Sn has been done recently in Ref. \cite{Litvinova2015} where they significantly contributed to the location of dominant single-particle states. The impact of coupling nucleons to isospin-flip vibrations on the two-body response has however never been studied so far. 
In this context, spin-isospin excitations of nuclei, such as Gamow-Teller (GT) transitions, constitute a great test. In approaches based on the RMF, such excitations are basically fully determined by pion exchange. The coupling to spin-isospin phonons which resums pion-nucleon dynamics to infinite order, as illustrated in Fig. \ref{f:exp-phon}-b), can therefore be expected to be important. In this letter we implement for the first time the coupling of CE vibrations to single nucleons in the description of GT transitions in doubly-magic neutron-rich nuclei. We investigate their impact on the quenching of the strength distributions and on the description of the low-energy states that determine beta-decay half-lives.

\section{Formalism}
The dynamics of an atomic nucleus in a weak external field $\hat F$ can be characterized by the transition strength distribution
\begin{eqnarray}
S(E) &=& \sum_N |\braket{\Psi_N|\hat F|\Psi_i}|^2 \delta (E - \Omega_N) \; ,
\label{eq:strength}
\end{eqnarray}
where $\ket{\Psi_i}$ and $\ket{\Psi_N}$ denote the nuclear ground and excited states respectively, and $\Omega_N = E_N - E_i$ are the corresponding excitation energies.
When $\hat F$ is a one-body charge-changing external field (e.g. containing the isospin-lowering operator $\tau_-$ transforming a neutron to a proton) 
the transition strength can be obtained from the two-body propagator, or response function, in the particle-hole proton-neutron channel $R_{pn'np'}(\omega)$ as, 
\begin{eqnarray}
S(E) &=& -\frac{1}{\pi} \lim_{\Delta\rightarrow 0^+} \mbox{Im} \sum_{pnp'n'} F^\dagger_{np} R_{pn'np'}(\omega) F_{p'n'} \; ,  
\label{eq:strength2}
\end{eqnarray}
where $p$ ($n$) denote proton (neutron) single-particle states, and $\omega = E + i \Delta$ is the complex energy variable. 
The response function $R_{pn'np'}(\omega)$ is obtained in the framework of the linear response theory, by solving the Bethe-Salpeter equation (BSE) \cite{BSE}. 
The effective interaction that enters this equation is determined consistently as the functional derivative of the nucleonic self-energy with respect to the one-body propagator \cite{Robin2016}. 
In this work, the static part of the self-energy is given by the RMF, while the dynamical part is obtained in the PVC framework, which accounts for virtual emission and re-absorption of nuclear vibrations by single nucleons. In the time-blocking approximation (TBA) \cite{Tselyaev1989,Litvinova2008} the final BSE coupled to a good angular momentum $J$ reads
\begin{eqnarray}
&&R_{(pn'np')}^{(J)} (\omega) = \widetilde{R}^{(0)(J) }_{(pn'np')} (\omega) + \nonumber \\
&&  \sum_{(p_1n_1p_2n_2)} \widetilde{R}^{(0)(J)}_{(pn_1np_1)} (\omega) W_{(p_1n_2n_1p_2)}^{(J)}  (\omega) R_{(p_2n'n_2p')}^{(J)}  (\omega) \; . 
\label{eq:BSE_pn}
\end{eqnarray}
In Eq. (\ref{eq:BSE_pn}) $\widetilde{R}^{(0)(J)}(\omega)$ is the RMF particle-hole proton-neutron propagator, and $W^{(J)}(\omega)$ denotes the two-body effective interaction. The latter is given by the sum of the static meson-exchange interaction in the isovector channel and the energy-dependent amplitude $\Phi(\omega)$ containing the effect of PVC:
\begin{eqnarray}
W^{(J)}(\omega) = \widetilde V_\rho^{(J)}  + \widetilde V_\pi^{(J)} + \widetilde V_{\delta_{\pi}}^{g'(J)} +  \Phi^{(J)}(\omega) \; .
\label{eq:int}
\end{eqnarray}
\noindent In Eq. (\ref{eq:int}) $\widetilde V_\rho $ and $\widetilde V_\pi$ are the finite range rho-meson and pion exchange interaction respectively, while $\widetilde V_{\delta_{\pi}}^{g'}$ denotes the zero-range Landau-Migdal term that accounts for short-range correlations \cite{Bouyssy1987}. Here we take the associated parameter $g'=0.6$, as the exchange interaction (Fock term) is not treated explicitly \cite{Liang2008}. 
Considering only the static interaction in Eq. (\ref{eq:int}), one gets back the proton-neutron relativistic RPA (pn-RRPA) \cite{Kurasawa2003}. 
\begin{figure}[t]
\centering{\includegraphics[width=\columnwidth] {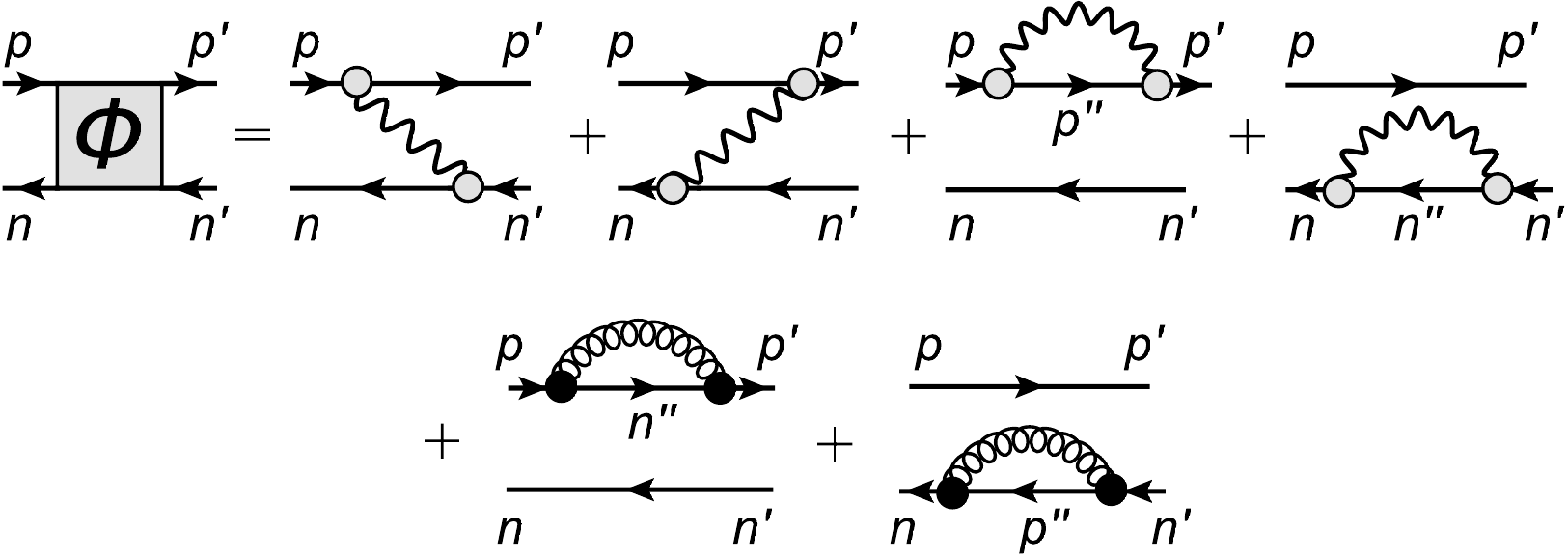}}  \hfill % 
\caption{PVC interaction with coupling to neutral (wiggly lines) and charge-exchange (springs) phonons. }
 \label{f:QVC_IVphon} 
\end{figure}
The PVC interaction $\Phi(\omega)$ introduces 1p1h$\otimes$phonon configurations and is responsible for damping of the transition strength beyond pn-RRPA. 
Labeling neutral and CE phonons by their excitation energies and quantum numbers $\{ \mu=(\Omega_\mu,J_\mu,M_\mu,\pi_\mu,T_z^\mu=0) \}$ and $\{ \lambda=(\Omega_\lambda,J_\lambda,M_\lambda,\pi_\lambda,T_z^\lambda=\pm 1) \}$, respectively, the PVC interaction reads
\begin{eqnarray}
\Phi^{(J)}_{(pn'np')}(\omega) =  \Phi^{\{\mu\} \, (J)}_{(pn'np')}(\omega) + \Phi^{\{\lambda\} \, (J)}_{(pn'np')}(\omega) \; ,
\label{e:Phi_IS}
\end{eqnarray}
and is shown in Fig. \ref{f:QVC_IVphon} in terms of Feynman diagrams. $\Phi^{\{\mu\}} (\omega)$ is the interaction induced by neutral phonons and its analytical expression can be found in Ref. \cite{Robin2016}. $\Phi^{\{\lambda\} }(\omega)$ is the new interaction induced by CE phonons. Due to charge-conservation, $\Phi^{\{\lambda\}} (\omega)$ only contains self-energy insertions, which take the same form as for neutral phonons, and no phonon-exchange term. We see from Fig.~\ref{f:QVC_IVphon} that these new self-energy terms involve proton-neutron particle-particle elements of the PVC vertex, and thus can be interpreted as a (virtual) energy-dependent proton-neutron pairing interaction in doubly-magic nuclei. In the following, the extension of the pn-RRPA including PVC effects in the TBA is referred to as pn-RTBA. \\ 

\noindent We emphasize that we do not apply any "subtraction procedure" \cite{Tselyaev2013}.
This procedure, which consists in subtracting $\Phi(0)$ from Eq. (\ref{eq:int}), had been introduced to avoid double counting of PVC effects that are implicitly contained in the mesons parameters.
However, in the case of GT transitions, the pion provides almost the whole contribution to the meson-exchange interaction, and, as it does not contribute to the RMF, is considered here with the free-space coupling \cite{Robin2016}. 
In this sense no double counting should occur when including PVC mechanism. If the Fock term, and therefore the pion, was included in the mean field, one would have to re-instore a subtraction method. However the subtraction in the CE channel would be problematic as poles can appear at zero energy in nuclei with isospin asymmetry, which forbids the subtraction of $\Phi(0)$. 

\section{Gamow-Teller transitions, quenching of the strength and beta-decay half-lives}
\begin{figure}[h]
\centering
{\includegraphics[width=\columnwidth] {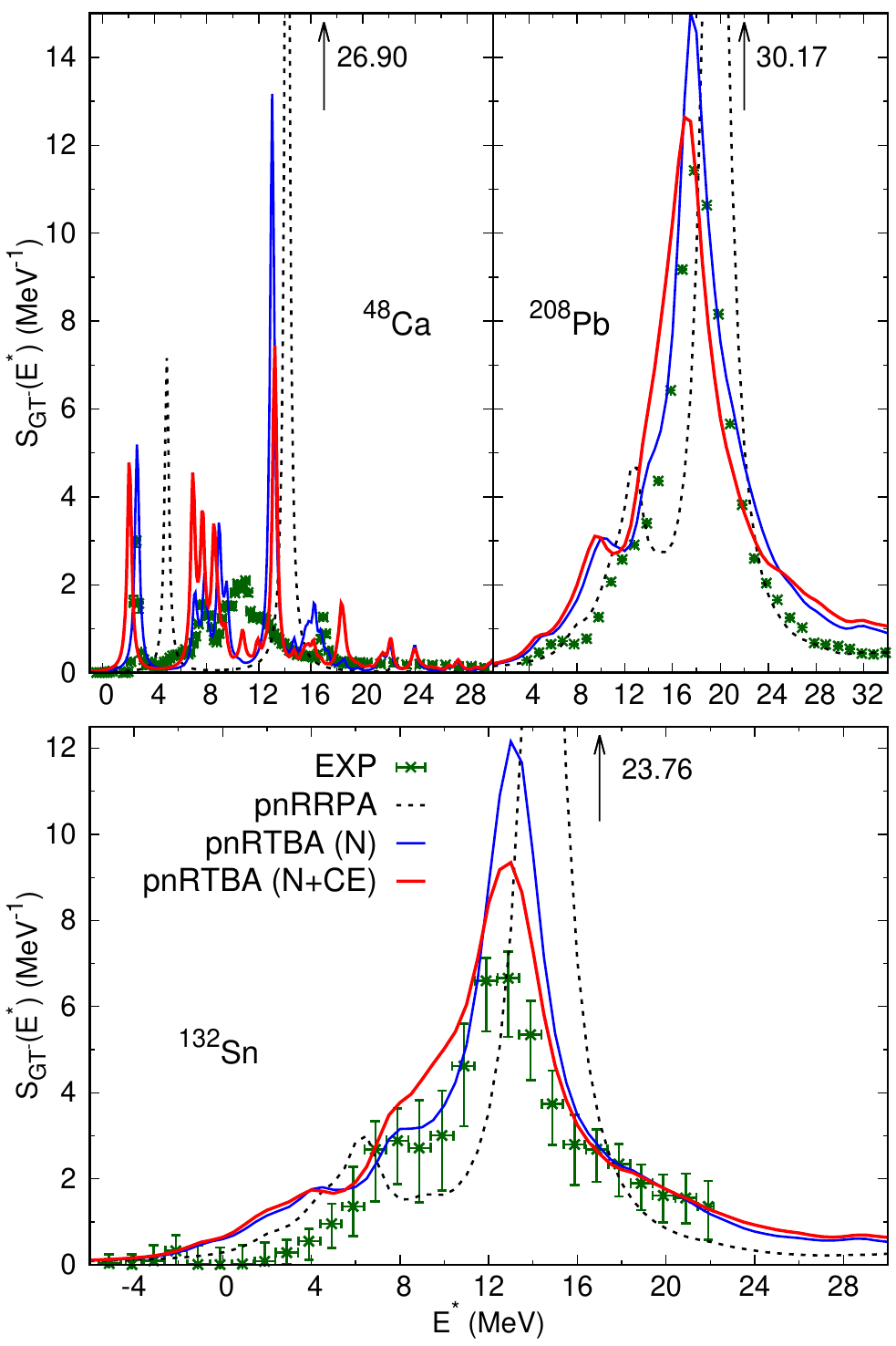}}  \hfill 
\caption{GT$^-$ strength distributions in $^{48}$Ca, $^{132}$Sn and $^{208}$Pb. The dashed black, full blue and full red curves show the results obtained within pn-RRPA, pn-RTBA with coupling to neutral (N) phonons, and pn-RTBA with coupling to neutral and CE phonons, respectively. The green points show the available experimental data \cite{Yako2009,Wakasa2012,Sasano18,Yasuda2018}.}
 \label{f:strength} 
\end{figure}
We apply the above formalism to the GT$^-$ response of doubly-magic neutron-rich nuclei. 
The corresponding field reads $\hat F = \sum_{i=1}^A {\boldsymbol \Sigma}_{(i)} \tau_-^{(i)}$, where $\boldsymbol\Sigma$ is the relativistic spin operator.
We solve Eq. (\ref{eq:BSE_pn}) using the following numerical scheme:  (\textit{i}) A RMF calculation is done using NL3 parametrization \cite{NL3} of the meson-nucleon Lagrangian.  (\textit{ii})  The spectrum of phonons that are coupled to nucleons is calculated within the RRPA and pn-RRPA. 
We include neutral phonons $\{ \mu \}$ with $J^{\pi}_{\mu} = 2^+,3^-,4^+,5^-,6^+$ as these were found to be sufficient in Refs. \cite{Robin2016,Yu1997}. CE phonons $\{ \lambda \}$ with $J_{\lambda}^{\pi} = 0^\pm,1^\pm,2^\pm,3^\pm,4^\pm,5^\pm,6^\pm,7^\pm$ are taken into account and their separate contribution is investigated in the following. The phonon spectrum is further truncated to keep those with excitation energies below 20 MeV and realizing at least $5\%$ of the highest transition probability for a given multipole.
(\textit{iii}) We solve the BSE (\ref{eq:BSE_pn}) for the proton-neutron response function with $J^\pi = 1^+$. 
PVC effects (i.e. 1p-1h $\otimes$ phonon configurations) are included in an energy window of 30 MeV around the Fermi level, which is the energy region of interest. \\ \\
In Fig. \ref{f:strength} we show the resulting GT strength distributions in $^{48}$Ca, $^{132}$Sn and $^{208}$Pb compared to the available experimental data \cite{Yako2009,Wakasa2012,Sasano18,Yasuda2018}. 
Taking a finite value of the smearing parameter $\Delta$ in Eq. (\ref{eq:strength2}) allows us to simulate the effect of continuum and configurations beyond 1p-1h $\otimes$ phonon that are not explicitly included in the present work.
In order to reproduce the experimental energy resolution we then take $\Delta=200$ keV in $^{48}$Ca and $\Delta=1$ MeV in $^{132}$Sn and $^{208}$Pb. The excitation energies on the x-axis are shown with respect to the parent ground-state, except for $^{48}$Ca that we shifted by the parent-daughter binding-energy difference calculated in the RMF, to compare to the data that is given with respect to the daughter ground state. 
The distributions obtained at the pn-RRPA level (fully neglecting the PVC interaction in Eq. (\ref{eq:int})) are in dashed black. In plain blue and red, are the strength functions obtained considering the coupling of nucleons to neutral phonons only, and to both neutral and CE phonons, respectively. 
As noted in Refs. \cite{Litvinova2014,Robin2016}, the coupling to neutral vibrations provides an important fragmentation of the pn-RRPA states and spreading of the strength towards both low- and high-energy regions.
In addition, the coupling to CE phonons introduces further modification of the strength with noticeable impact on the quenching of the giant GT resonance (GTR) which appears particularly important in $^{48}$Ca and $^{132}$Sn. %
In $^{48}$Ca the main GTR peak around 13 MeV is reduced by $\sim 43 \%$, which is of the same order as the effect produced by neutral phonons on the pn-RRPA resonance.
As seen on the data, a well-defined state appears above the GTR, due to the coupling to the $0^+$ CE phonon. 
In $^{132}$Sn the width of the GTR is increased by the inclusion of CE phonons and the value of the GTR peak is reduced by $\sim 24\%$, which is counter-balanced by the appearance of the "shoulder" structure right below the GTR. These modifications make the trend of the distribution in very nice accordance with the recent measurement done at RIKEN \cite{Sasano18,Yasuda2018}. In $^{208}$Pb, the shape of the experimental distribution is very well reproduced by the inclusion of the CE phonons with almost no necessary quenching. \\ 

\begin{figure}[h!]
\centering
{\includegraphics[width=\columnwidth] {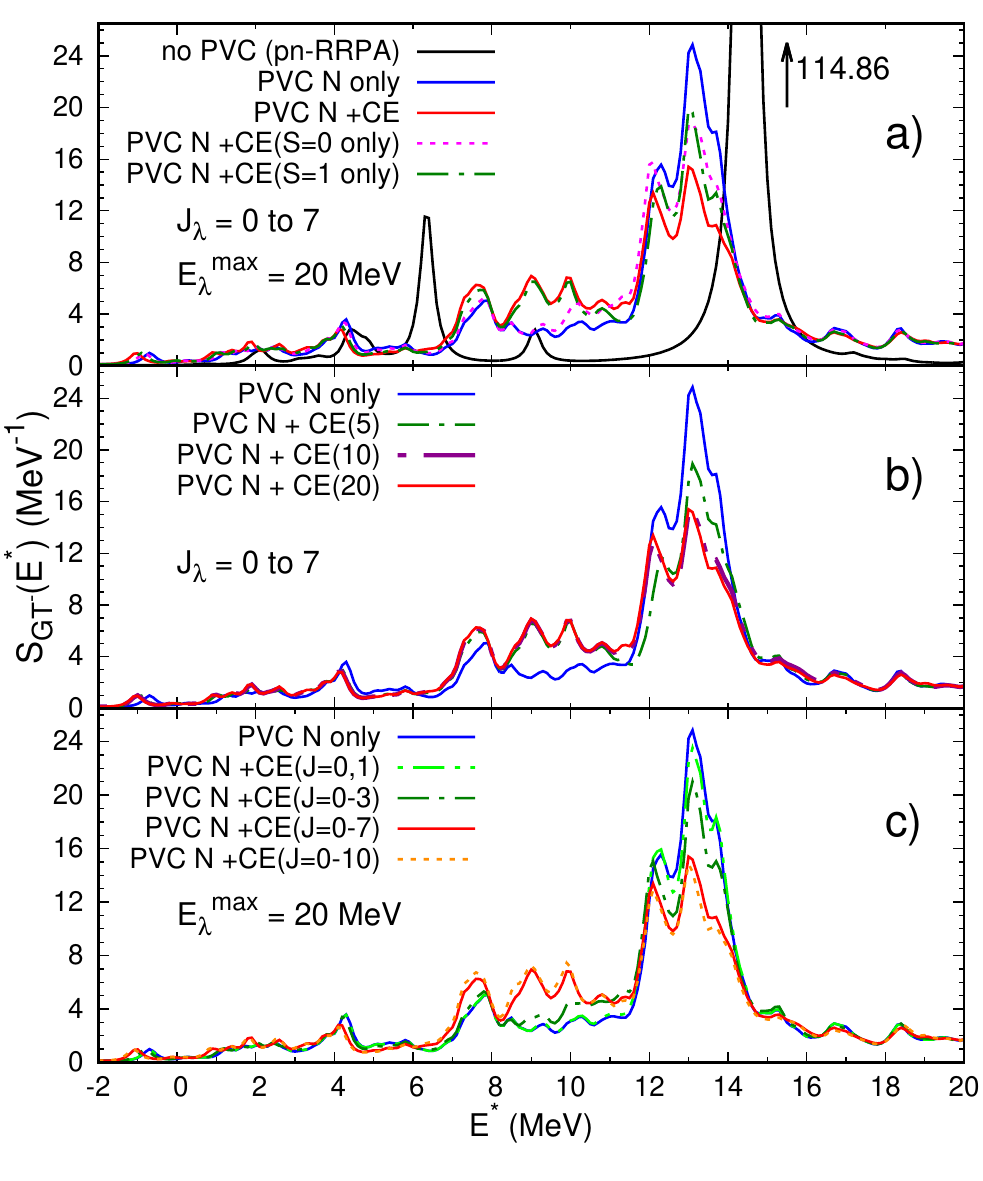}}  \hfill % 
\caption{GT$^-$ strength in $^{132}$Sn with $\Delta=200$ keV, for different truncations of the spectrum of CE phonons $\{\lambda\}$. See text for details.}
 \label{f:132Sn-1} 
\end{figure}

In order to disentangle the contributions of different CE phonons we show in Fig. \ref{f:132Sn-1} the GT distributions in $^{132}$Sn obtained for different truncations of the phonon spectrum $\{\lambda\}$, using a smearing $\Delta=200$ keV.
We show in Fig. \ref{f:132Sn-1} a) the separate contributions of CE phonons without and with spin exchange ($S^\lambda$=0,1). 
While the modification of the GTR peak around 13 MeV is due to both types of vibrations, the shoulder structure below is mainly caused by spin-isospin modes ($S^\lambda$=1). 
This emphasizes the role of dynamical contributions of the pion to GT modes, and thus the importance of non-linear effects.
In Fig. \ref{f:132Sn-1} b) we show the GT distribution obtained when varying the truncation energy of the spectrum of CE phonons. 
The maximal phonon energy is given in parenthesis, in MeV.
The coupling to CE vibrations below 5 MeV actually gives the most important contribution to the shoulder around 9 MeV. The quenching of the GTR appears gradually when including phonons up to 10 MeV. CE vibrations with excitation energies from 10 to 20 MeV introduce only very little change.
Finally Fig. \ref{f:132Sn-1} c) shows the transition strength for CE phonons with different angular momentum and parities. 
CE phonons with $J_\lambda=0,1$ give almost no contribution in this nucleus while the ones with $J_\lambda=2,3,4,5,7$ appear to couple non-negligibly to nucleons. 
The contribution of phonons with $J_\lambda=8,9,10$ remains smaller. \\
\begin{figure}[h]
\centering
{\includegraphics[width=\columnwidth] {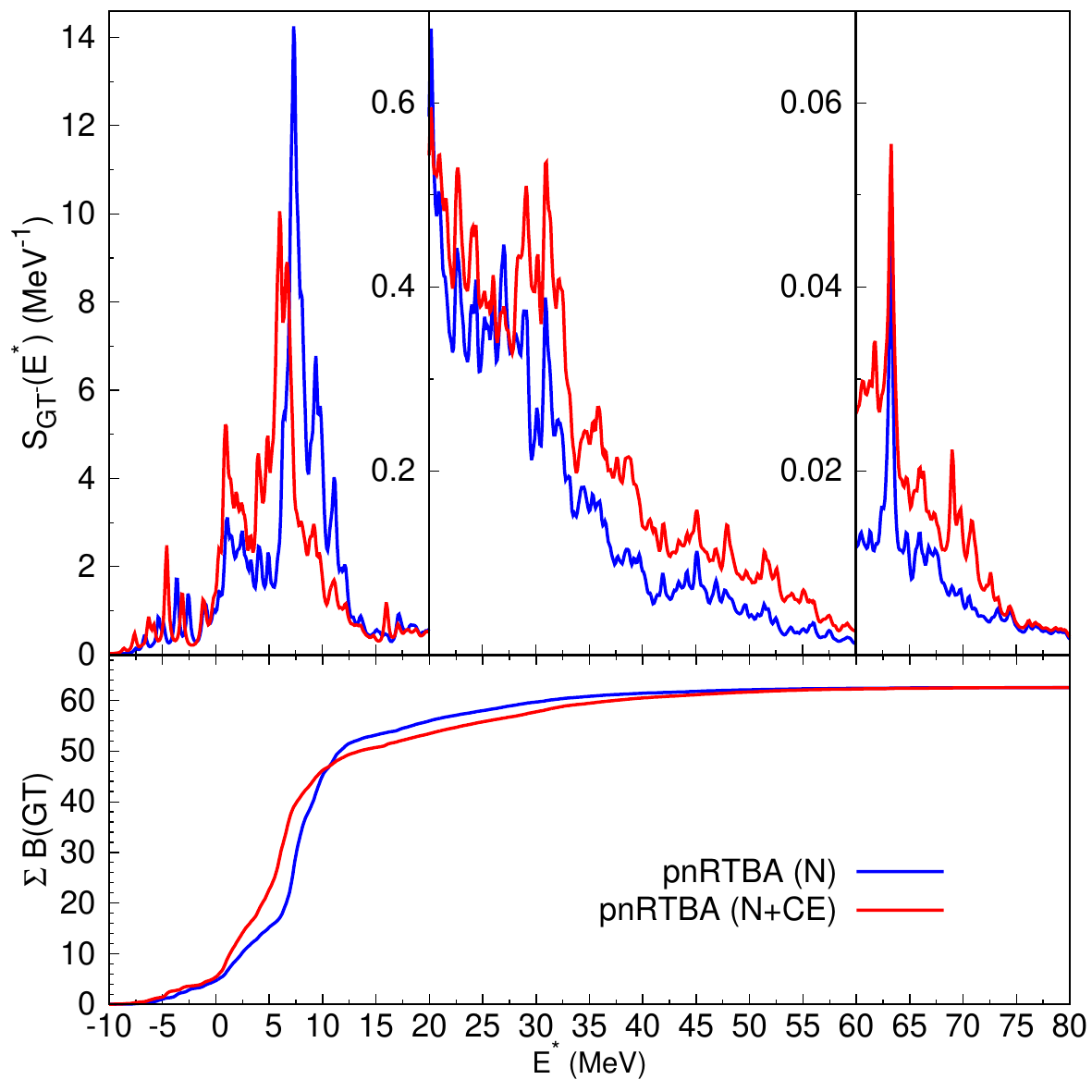}}  \hfill % 
\caption{GT$^-$ strength distribution in $^{78}$Ni (top) and cumulative integrated strength (bottom), with $\Delta = 200$ keV.}
 \label{f:78Ni} 
\end{figure}

It is well known that charge-exchange experiments only observe about $\sim 60-70\%$ of the Ikeda sum rule up to the GTR region. This suggests that an important fraction of the transitions appears at higher excitation energies. Several mechanisms have been proposed to explain such phenomenon, including the possible excitation of a Delta resonance \cite{Kurasawa2003} (not considered in the present work), and the coupling of 1p-1h proton-neutron excitations to higher-order configurations such as 2p-2h. In order to investigate the latter mechanism, we must include complex configurations in a model space as large as possible.
In $^{78}$Ni, we can increase the PVC energy window, in which 1p-1h$\otimes$phonon excitations are included, up to $100$ MeV that is the same cut-off used to select the 1p-1h pairs in pn-RRPA. 
We show in Fig. \ref{f:78Ni} the resulting distribution and corresponding cumulative strength. When coupling nucleons to neutral phonons, we find about $10 \%$ of the strength above 20 MeV (compared to $2\%$ in pn-RRPA), while this number grows to $15 \%$ when including the coupling to CE phonons. This study therefore tends to agree with the fact that the experimental missing strength lies in the contribution of many high-energy transitions with very small strength, and, in this context, demonstrates the importance of the new phonons.
We note that the saturated GT$^-$ strength does not add up to 66 units, due to a) the non-zero strength in the GT$^+$ branch, b) the contribution of transitions to the Dirac sea \cite{Paar2004}. The overall Ikeda sum rule is fulfilled in both pn-RRPA and pn-RTBA. \\

Since the coupling to CE phonons generally causes a modification of the very low-energy strength, we investigate the impact of such effects on beta-decay half-lives. These are calculated as in Refs. \cite{Niu2015,Robin2016}, using here the bare value of the weak axial coupling constant $g_A \simeq 1.27$ \cite{gA}.
The results for $^{78}$Ni and $^{132}$Sn are shown in Table \ref{t:h-lives}.
\begin{table}[h]
 \caption{Beta-decay half-lives of $^{78}$Ni and $^{132}$Sn, in seconds.}
  \centering
\begin{tabular}{ccccc}
Nucleus & EXP\cite{Hosmer2005,nndc} &RRPA & RTBA (N) & RTBA (N + CE) \\
\hline 
$^{78}$Ni & $0.110^{+0.210}_{-0.05} $ &   1.8806        &	0.0731 &	0.0395 \\ 
\hline
$^{132}$Sn &$39.7^{+0.8}_{-0.8}$ 	&	1626.729  &	36.427 &	16.592  \\
\hline
\end{tabular}
\label{t:h-lives} 
\end{table}
The coupling between nucleons and neutral vibrations causes a decrease of both half-lives due to the redistribution of the strength, leading to a better agreement with the experimental value, compared to the pn-RRPA results. This was observed in Ref. \cite{Robin2016} where the chain of Ni isotopes was studied. 
The coupling to CE phonons produces further modification and shift of the low-energy strength, yielding an additional decrease of the half-lives, that are now slightly underestimated. The experimental order of magnitude is however reproduced in $^{132}$Sn.
We remind that meson-exchange currents are not considered here, while they are known to induce quenching of beta-decay rates. 
Moreover, no ground-state correlations induced by PVC are included, while they typically correct for a too strong shift of the low-energy strength \cite{Robin2018}.
The coupling to CE phonons could then provide a mechanism to lower the half-lives.

\section{Summary and conclusion}
We have advanced the formalism of the nuclear isospin-flip response theory to include polarization effects induced by charge-exchange phonons, which were not considered previously. Thereby, we have taken into account dynamical pion and rho-meson exchange in both the single-particle and two-body motion up to infinite order, which represents a conceptual advancement of the isovector sector of Quantum Hadrodynamics.
The extended formalism is implemented for investigation of the Gamow-Teller response in neutron-rich doubly-magic nuclei. It is shown that this new mechanism can explain a sizable part of the quenching observed in charge-exchange experiments
and can contribute significantly to the details of the low-energy strength that predicts beta-decay half-lives, with important astrophysical implications, especially around the r-process waiting point nuclei $^{78}$Ni and $^{132}$Sn. 
The obtained results emphasize the utmost importance of non-linear effects for an accurate microscopic theory of nuclear charge-exchange and weak processes. 

\section*{Acknowledgments}
Discussions with G. Col\`o are gratefully acknowledged. This work was supported by the Institute for Nuclear Theory under US-DOE Grant DE-FG02-00ER41132, by JINA-CEE under US-NSF Grant PHY-1430152, and by US-NSF grant PHY-1404343 and US-NSF CAREER grant PHY-1654379.

\bibliography{mybibfile}

\end{document}